\documentstyle[epsf,twoside,oldlfont]{ima}
\input amssym.def
\input amssym

\begin{document}
\contribution{aizenman}
\markboth{MICHAEL AIZENMAN}{INCIPIENT SPANNING CLUSTERS}

\newtheorem{thm}{Theorem}
\newtheorem{lm}[thm]{Lemma}
\newtheorem{df}{Definition}
\def\H{{\mathcal H}}
\def\be{\begin{equation}}
\def\ee{\end{equation}}
\def\bea{\begin{eqnarray}}
\def\eea{\end{eqnarray}}
\def\remark{ \noindent {\bf Remark:}  }
\def\proof{ \noindent {\bf Proof:}  } 
\def\P{Prob_{p_c}}   
\def\E{{\mathbf E}}  
\def\I{\mathrm I}    
\def\C{{\mathcal C}}   
\def\atoo{\parbox[t]{.4in}
        {$\longrightarrow \\ {\scriptstyle \alpha \to 0}$}}
                                                        
\newcommand{\eq}[1]{eq.~(\ref{#1})} 

\def\lg{\stackrel{\scriptstyle <}{_{_{\scriptstyle >} }} }
\def\L{ {\mathcal L} }     
  

  \begin{figure}[t]{ \footnotesize   
 For: ``Mathematics of Materials:  Percolation and Composites'', Editors: 
 \\ \hfill 
  \quad K.M. Golden, G.R. Grimmett,
  R.D. James, G.W. Milton, and P.N. Sen.   \\ \hfill  
  \quad The IMA Volumes in Mathematics  and its Applications 
  (Springer-Verlag, 1997).}
  \end{figure} 

\title{SCALING LIMIT FOR THE INCIPIENT SPANNING CLUSTERS}
\author{Michael Aizenman\thanks{ 
Departments of Physics and Mathematics,
Jadwin Hall, Princeton University,
Princeton, NJ 08544-0708. } } 
\maketitle
\begin{abstract}
Scaling limits of critical percolation models show major 
differences between low and high dimensional models.   
The article discusses the formulation of the continuum limit
for the former case.  A mathematical framework is proposed
for the direct description of the limiting continuum theory.
The resulting structure is expected to exhibit 
strict conformal invariance, and facilitate the mathematical 
discussion of questions related to universality of critical
behavior, conformal invariance, and some relations with a number of
field theories.
\end{abstract}


{\AMSMOS  82B43, 82B27, 60D05, 82-02   
\endAMSMOS}

\renewcommand{\thefigure}{\arabic{figure}}
\renewcommand{\thetable}{\arabic{table}}

  \renewcommand{\baselinestretch}{1}
  {
  \renewcommand{\thefootnote}{}
  \footnotetext{
  {\copyright \ Copyrights rest with the author.  Faithful 
  reproduction for non-commercial purpose is permitted.}}

\section{Introduction} 
Incipient percolation clusters have attracted attention as objects
of interesting physical and mathematical properties, and potential 
for applications.   An example of a setup in which they play a role 
is an array of conducting elements, placed at random in an
insulating medium, with the density  adjusted to be close
to the percolation  threshold.  In such arrays
the current is channelled through fractal-like sets.  
The concentration of the current, or stress/strain in other similar
setups, may result in high amplification of non-linear effects.
The phenomenon is of technological interest, and plays a role in 
high-contrast composite materials and non-linear composites,  
utilized in thermistors and other devices \cite{BI,composits,binaries}. 
Studies of the relevant random geometry have yielded interesting 
geometric concepts such as the celebrated (but often misunderstood)
Incipient Infinite Cluster (IIC).
The topic was reviewed from a physics perspective in an article 
(Stanley \cite{Sta_rev}) which appeared in Volume 8 of this series,  
in the proceedings of a workshop held at IMA in 1986.  

It is somewhat surprising that percolation threshold phenomena are 
still a source of delightful and new observations, since the subject 
seemed to be reaching its maturity already ten years ago.
Nevertheless, the subject has recently enjoyed renewed attention;
in part because it was realized that some entrenched notions
need correction (ref. [5-13]),
and in part because it was realized that the scaling (continuum) 
limit has interesting properties, e.g., conformal invariance, 
and its construction presents an interesting mathematical 
challenge [14-16, 10].

This article focuses on issues related to the 
{\em scaling limit\/} of the Incipient Spanning Clusters (ISC).  
We discuss a mathematical framework for the direct 
description of the limiting continuum theory, which is applicable
to models below the upper--critical dimension.  
Within it, we encounter some interesting fractal structures 
and questions related to: universality of the critical behavior, 
conformal invariance and relations with certain field theories.
The purpose is to describe some recent developments
and point directions for possibly interesting progress. 

\section{The incipient infinite cluster and the 
incipient spanning clusters}

A good starting point for the discussion of the scaling limit
is the conceptual  difference between two related terms: the
Incipient Infinite Cluster (IIC), and the 
Incipient Spanning Clusters (ISC) 
(interpreted here as in ref.~\cite{Aiz96}).

The Incipient Infinite Cluster (IIC) is a thought--provoking notion 
which has  often been used in the discussion of different 
aspects of critical percolation phenomena (\cite{Sta_rev}).  
It has provided a useful and stimulating concept, 
but one whose different uses have led to some
confusion and misinterpretation of theoretical predictions,
in particular concerning the question of uniqueness 
(see the Stauffer paradox discussed in \cite{Aiz96}).  
Consequently, it was recently proposed 
to differentiate between the two related notions
mentioned above: the IIC and the ISC.
The distinction is most apparent in the limits in which 
the two are described by different mathematical entities. 

The Incipient Infinite Cluster is a random infinite cluster
(or, more completely, a random environment exhibiting an infinite
cluster), constructed by  a limiting process which 
provides the ``local'' picture of the large clusters seen at 
the percolation threshold --- viewed
from the perspective of one of their sites.   
This notion covers three alternative paths which 
have been explored towards the construction of the IIC: 
\begin{itemize}
\item[i.]
Condition on the origin being connected
distance $L$ away, and let \mbox{$L\to \infty$}.
\item[ii.]  Raise the percolation density above the percolation threshold,
condition on the origin belonging to an in infinite cluster, and 
then let $p \searrow p_c$ (where the percolation 
density presumably vanishes).  
\item[iii.] Generate the probability distribution for a random cluster
(or a random environment) by centering a typical random 
configuration relative to one of the sites on its spanning clusters 
[or just large clusters].  The sampling is to be done
with equal weights over all the sites 
connected to the boundary of $[0,L]^d$ (and then let $L\to \infty$), 
possibly with a corrective exclusion of a boundary zone.  
\end{itemize}
The first two procedures were pointed out by Kesten, who proved their 
convergence and equivalence in $d=2$ dimensions  \cite{K_IIC};
the third is more convenient for numerical studies, and was used in  
\cite{HASM}.
Either way, one may see that the mathematical construction
covers the microscopic view of the incipient infinite cluster from the
view of one of the rare sites which at $p=p_c$ are connected very
far -- on the microscopic scale.   
(A somewhat different conception of the IIC is presented in ref~\cite{CCD}. 
Presumably all will agree in the further scaling/continuum limit).  

The Incipient Spanning Clusters, on the other hand, are naturally 
viewed on the ``macroscopic'' scale.  
They are simply the large clusters which reach across the finite 
sample (in the above example), and connect opposite boundary 
segments, as indicated in Figure~\ref{ISC}.  
In order to see the entire collection of spanning clusters, we need to 
keep track of events on the scale of the sample, 
the relevant limit being: {\em lattice spacing}
($\alpha$) $\to 0$.  
 
\begin{figure}[ht]
    \begin{center}
    \leavevmode
    \epsfbox{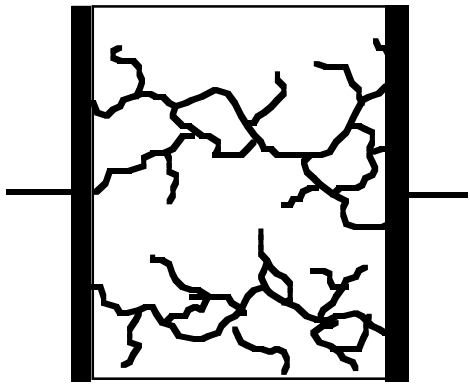}
  \caption{Incipient Spanning Clusters --
   a schematic depiction of the the macroscopic view.}
\label{ISC}
\end{center}
\end{figure}

A particular difference which stands out, and one
which has caused extensive confusion and discussions,  is 
that the the IIC typically shows a single dominant (infinite) 
cluster, whereas the ISC quite generally exhibit multiple clusters 
of comparable size, \cite{Aiz96}.  
(A more complete discussion of the related {\em Stauffer paradox} 
is found in ref~\cite{Aiz96}).
   
The continuum limit enables a natural discussion of the enhanced 
symmetry. The highest symmetry emerges at the critical point,
for which there is strong evidence of conformal invariance.
Considerations related to 2D conformal fields have
led to proposals for differential equations which determine some
of the properties of the critical measure (\cite{Car,LPS}).
Thus, the continuum limit of the spanning clusters may
remind one of the Brownian motion: an object arising from physics, with
fascinating mathematical properties, high degree of symmetry, and
relation to interesting differential equations.

While we focus here on ICS, similar considerations
can be applied to the entire collection of the
connected clusters which are visible on the macroscopic scale.  
The entire ensemble's scaling limit has been termed the 
{\em percolation web\/} \cite{Aiz96}. 

\bigskip
\section{The microscopic view; three convenient models} 

There is a variety of situations in which 
conducting elements are placed at random in an insulating medium.
These elements may conduct electric current, or may
serve as passages for a liquid seeping through a solid.
The individual resistors, or cracks, are visible on
the {\bf Microscopic Scale}.  Their relative density
is our (dimensionless) control parameter.
Following are three convenient models, which offer different 
advantages as possible starting points for the construction of a 
(common?) continuum limit. 

\addtocounter{figure}{1}
\begin{figure}[ht]
    \begin{center}
    \leavevmode
    \epsfbox{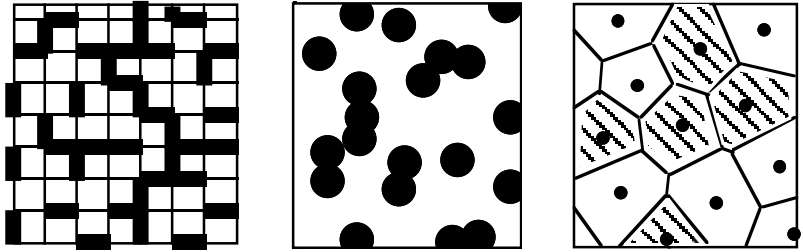}
  \caption{The microscopic view;
   different possibilities for the short--distance structure.}
\label{micro}
\end{center}
\end{figure}

\vspace{\abovedisplayskip}
\noindent {\bf Bond-percolation.}   The model is
formulated on the lattice $\alpha Z^d$.  The randomness is 
associated with the bonds (pairs of neighboring lattice sites),
which are occupied at random, independently with probability 
$0 \le p \le 1$ (the control parameter).  The occupied
bonds are regarded as connecting.  They may also represent
conductors of conductance $\sigma > 0$.

When a bond is not occupied, its dual obstructing object
can be regarded as realized.  That leads
to the self-duality of bond percolation on $Z^2$, which 
is helpful (\cite{k_pc,K_scaling}).  
In three dimensions, the dual  model is that 
of random plaquettes, which may form encapsulating 
surfaces obstructing the connections (\cite{ACCFR}).

\vspace{\abovedisplayskip}
\noindent {\bf Droplet percolation.}  This models is formulated
over the continuum.
The conducting regions consist of randomly distributed d-dimensional
balls of radius $\alpha$ (Poisson process, with possible overlaps),
with density $\rho d^dx$ for the centers  of the conducting cells.  
The relevant dimensionless control parameter is proportional to 
the density of the conducting regime: 
\be
  p \ =\ \rho \ \alpha^d \  .
\label{eq:r-a}
\ee
 
The 2D bond model is self dual, while the droplet model is rotation
invariant.  The following model exhibits both
features (a fact noticed
independently  also by Benjamini and Schramm (\cite{BS}),
and it, therefore, is our favored starting point for the
construction of the purportedly common\ scaling limit. 

\vspace{\abovedisplayskip}
\noindent {\bf Voronoi-tessellation percolation.}  
Starting from a randomly generated configuration of points in $R^d$,
described by a Poisson process with density $\rho d^dx$,
the plane is partitioned into the cells of the corresponding Voronoi
tessellation.  The cells are conducting, or not, independently with
probabilities $\{p,1-p\}$.   Alternatively stated, two discrete random 
sets ($A,B\subset R^d$) are
generated with Poisson densities $p \rho d^dx$ and $(1-p) \rho d^dx$,
and the conducting regime consists of those sites of the continuum
which are closer to A than to B.  
The short-distance scale in this model, $\alpha$,  is related
to  $\rho$ as in eq.~\ref{eq:r-a} (it is of the order of the mean 
diameter of the Voronoi cells). In two dimensions the model
is self-dual, and the critical value for $p$ is $p_c=1/2$.

An generalization we shall mention in Section~\ref{Conf-inv} consists 
of models with a density profile of the form $\rho_t(x) d^dx$ with
$\rho_t(x) = t\cdot g(x)$, $g(x)$ continuous and positive.   
Letting \mbox{$t\to \infty$} we find 
a scaling limit for which the 
density profile shows persistent {\em inhomogeneity} on the 
macroscopic scale.  However, since the inhomogeneity corresponds to
just different rates of approach towards a common limit, we 
expect it to have no visible effect on the continuum limit
considered here. 

Planarity is a very helpful property
even without strict self duality.
Using it one can prove more for 2D models, at or near $p_c$ 
(\cite{K_scaling,Alex_RSW}), 
than what is known about other dimensions
$2<d<6$, and in particular about $d=3$.
 
\section{The macroscopic perspective}
   
The focus of our discussion is on the geometric 
features which are visible on the {\bf Macroscopic Scale}
in a systems whose short scale structure is any of the above. 
Correspondingly, we chose the scale for our discussion so that the object 
occupies a fixed continuum--scale region $\Lambda_1=[-1,1]^d$,  
or more generally $\Lambda_{\ell} = [-\ell,\ell]\times [-1,1]^{d-1}$, 
and we let the short distance scale be $\alpha << 1$, eventually taken to $0$.  

When the sample is placed between two conducting  plates, with 
different electric potentials applied to the two opposite faces:  
\be
\partial \Lambda_{-}=\{\underline{x}\in \partial \Lambda \ : \  
x_1=-\ell \}\ , \
\ \partial \Lambda_{+}=\{\underline{x}\in \partial \Lambda \ : \ x_1=\ell \}
\ , 
\ee
we would naturally be interested in  the spanning clusters, which are
the (maximal) connected clusters linking $\partial \Lambda_{-}$ with 
$\partial \Lambda_{+}$.

While the changes on the {\em microscopic} scale are gradual, 
at the percolation threshold a drastic 
transition is observed on the {\em macroscopic} scale, where the
following is seen  
with probability which tends to $1$  as $\alpha \to 0$
(for suitably chosen constants $Const.$):
\bigskip
\begin{description} 
\item [$p<p_c \Rightarrow$]
there are no spanning clusters; the diameters of the  
connected clusters do not exceed $Const. \ \alpha |\log \alpha|$.
\bigskip
\item [$p>p_c \Rightarrow$] 
there is a unique spanning cluster, which 
covers the region \\
``densely'': its spherical voids are all
smaller  than $Const. \ \alpha |\log \alpha|$.  
\end{description}
\bigskip
\noindent At the critical point we find:
\bigskip

\noindent $p=p_c \Rightarrow$ \mbox{ } \\ 
\begin{itemize}
\item [i.]  the spanning probability does not vanish,
at least for $\ell < 1/2$ (the restriction is not needed for 
$2D$, otherwise it reflects just a 
limitation of the existent proof), \\
\item [ii.] for each $0<s<1$ --
\end{itemize}
\bea
Prob_{p_c, a}\left(
\begin{array}{l}
  \mbox{the inner region $[-s,s]^d$ is connected} \\
  \mbox{to the boundary $\partial \Lambda$}
\end{array} 
     \right) \ \ge \ C_d \ s^{(d-1)/2}   \nonumber \\
  \mbox{ }
\eea
\begin{itemize}
\item[] where the bound is {\em uniform in $\alpha$} (!), \\
\item [iii.]  there is positive probability for
more than one spanning clus-\\ ter in $\Lambda$.
\end{itemize}

\bigskip

\noindent {\bf Remarks:}  The proofs of the above assertions are based on a 
number of different results, some of which were presented most 
completely for the lattice -- rather than continuum, models.
The behavior at $p<p_c$ follows from the exponential decay of the 
two point function in the subcritical regime (\cite{AB,Men}), 
the bound on the voids in the percolating phase uses the coincidence
of the critical point with the limit of the slab/quadrant percolation
thresholds (\cite{BGN,GM}), as explained in \cite{Aiz96}.  The 
statements concerning $p=p_c$ in arbitrary dimensions are proven 
in \cite{Aiz96}.  The picture described there
fits well in the axiomatic description of the critical regime 
proposed in ref.~\cite{BCKS}.  The non-uniqueness of the spanning 
clusters in the two--dimensional case has been
familiar to those following the rigorous arguments since the work of
Russo \cite{R} and  Seymour and Welsh \cite{R,SW}, 
but it became appreciated in some 
of the physics community only rather recently \cite{Hu,Sen}.

Many (though not all 
\footnote[1]{The effective conductance is an example of a quantity
whose critical power law shows difference between some of the 
models listed above \cite{FHS}}) 
of the features of the scaling limits
of critical percolation are expected to be shared by systems
which differ on the microscopic scale.  
However, beyond the general description presented above, 
the behavior at $p=p_c$  exhibits interesting dimension dependence.

\section{Type I and type II critical models}

\addtocounter{figure}{2}
\begin{figure}[ht]
    \begin{center}
    \leavevmode
    \epsfbox{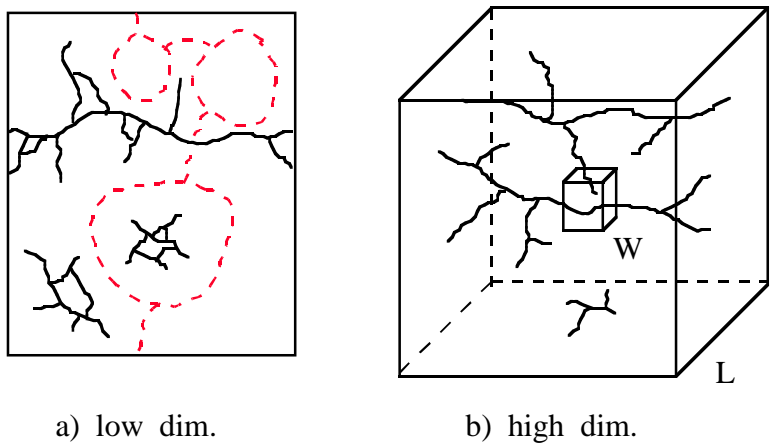}
  \caption{Two different types of critical behavior (schematically): 
  I) In $2D$ 
  one sees only a finite number of clusters of the size of the 
  volume, typically none of the clusters intersecting a given cube 
  reaches far beyond it -- on the corresponding scale, and in the 
  scaling limit the spanning clusters are nowhere dense. 
  II) In high dimensions  ($d>6 \ ?$) spanning clusters proliferate
  \/ $(O(L^{d-6}))$\/ and extend increasing distances. 
  While individual spanning clusters behave as $D=4$ dimensional 
  trees, their union becomes dense in the scaling limit.}
\label{low-high}
\end{center}
\end{figure}

The critical behavior is better understood 
in dimensions $d=2$, and $d > d_{u.c.}$ 
where $d_{u.c.}$ is the upper--critical 
dimension whose value for percolation is apparently $d_{u.c.}=6$ 
([33-37]).
It was, however, recently realized that two cases differ in ways
which have pronounced effects on the basic 
structure of the continuum limits.

In $d=2$ dimensions,  the  ``Russo--Seymour--Welsh theory'' 
\cite{R,SW} (conveniently summarized in \cite{G}) can be used to 
deduce that:
\begin{itemize}
\item [1.] A typical configuration has only finitely many 
spanning clusters -- in the sense that
\end{itemize}
\bea
Prob_{p_c,\alpha}
\left(\mbox{there are in $\Lambda$ $n$ distinct spanning clusters} \right) 
\ \le \ K(n) \ , \nonumber \\
\mbox{ }
\eea
uniformly in $ 0 < \alpha \le 1$, with $K(n) \to 0$ (for $n\to \infty$).  
More explicitly, 
$K(n)$  was recently shown to be bounded above $(C)$ and below $(C')$
by $A\ e^{-C[C`] \ n^2}$, \cite{Aiz96}.

\begin{itemize}
\item [2.] 
Any given internal site in $\Lambda$ is typically surrounded on an 
infinite sequence of scales ($2^{-n}$; $n=k_1,k_2,\ldots$) by 
``dual circuits'' which separate it 
from all the spanning clusters, and in the limit $\alpha \to 0$ from all 
clusters whose size remains visible on the macroscopic scale.  
Of course, exceptions to the rule are found along the spanning 
clusters, which form random ``fractal'' sets of lower 
dimension ($ < d = 2$).
\end{itemize}

In contrast, for dimensions $d > 6$ under an additional assumption,
that $\eta = 0$ (as predicted by physical theory \cite{HL} which has been 
supported by the partial rigorous results \cite{HS,Sl}), we find the 
following behavior  \cite{Aiz96}:
\begin{itemize}
\item [1.]
the number of spanning clusters grows, typically, as $a^{-(d-6)}$
(as predicted in ref.~\cite{coniglio})
\item [2.] 
the spanning probability tends to $1$ and,
furthermore, for any fixed open set  $B\subset \Lambda$ the 
probability that a spanning cluster intersects $B$ tends to $1$
\item [3.] 
the diameter of the maximal cluster intersecting a given open set,
of fixed  size on the macroscopic scale ($[-s,s]^d$), diverges as 
$\alpha \to 0$, 
typically being at least as large as  
$s^{(d-4)/2}\cdot \alpha^{-(d-6-o(1))/2}$.
\end{itemize} 
\noindent One could add that in the latter case the clusters have 
predominantly tree characteristics, and behave as $D=4$ dimensional 
objects, as was first suggested in ref.~\cite{AGK,AGNW}. 

The above examples, and the results presented below,
motivate the distinction which was made in ref.~\cite{Aiz96} 
between the following two types of critical behavior.

\noindent
{\bf Type I models:} The function
\bea
        \limsup_{\alpha \to 0} \ Prob_{p_c,\alpha}\left(
        \begin{array}{l}
                \mbox{the set $[-s,s]^d$ is connected}  \\
                \mbox{to the boundary of $[-1,1]^d$}
        \end{array}
           \right) \ \ \ & = &    \\
           \limsup_{\alpha \to 0} Prob_{p_c,\alpha}\left(
        \begin{array}{l}
                \mbox{the set $[-1,1]^d$ is connected}  \\
                \mbox{to the boundary of $[-1/s,1/s]^d$}
        \end{array}
           \right) \ & = & \tilde{h}(s)  
            \nonumber  
\label{5.c}
\eea
is strictly less than one, for some $ s<1$.  This implies
\mbox{$\lim_{s\to \infty} \tilde{h}(s)=0 $}, which means that there 
is no percolation in the scaling limit. 

\noindent{\bf Type II models:}
\begin{equation}
         Prob_{p_c,\alpha}\left(
        \begin{array}{l}
                \mbox{the set $[-s,s]^d$ is connected}  \\
                \mbox{to the boundary of $[-1,1]^d$}
        \end{array}
\right) \atoo 1 \ ,
\label{5.d}
\end{equation}
for any  $0 < s < 1$.

\noindent 
{\bf Remarks:\/} 1) Presumably all other behavior is ruled out
for the models considered here, but that was not proven. \\
2)  The function $ \tilde{h}(s) $ is obviously submultiplicative, 
i.e., satisfies $\tilde{h}(s_1 \cdot s_2) \ \le \ \tilde{h}(s_1) 
\tilde{h}(s_2)$ for $0 < s_1, s_2 < 1$.  Standard arguments imply
the existence of the limit, and its positivity for Type I systems:
\be
\lim_{s\searrow 0} \frac{\log \tilde{h}(s)}{\log s} 
\ = \ \lambda \ > \ 0 \ .
\label{lambda}
\ee
In other words, in Type I critical models
$\tilde{h}(s) \ = \ s^{\lambda +o(1)}$ for $s\to 0$, with 
some $\lambda > 0$.  In terms of the standard, though not yet fully proven
picture \cite{Sta_rev}, the exponent $\lambda$ is related to the Hausdorff 
dimension of the Incipient Spanning Cluster (or of the ISC), $d_w$, as:
\be
\lambda \ = \ d - d_w \ .
\ee

In this article we shall not discuss the scaling limit of models of 
of Type II (i.e., the case $d>d_{u.c.}$).  
Not that this would be uninteresting:
Hara and Slade proposed 
(as mentioned in \cite{DS}) that the limit for individual clusters
is related with the Integrated SuperBrownian Excursion process (ISE)
of Aldous \cite{Al}.   Furthermore, looking at the ensemble of all the 
macroscopic scale cluster we find that 
in a sense which still has to be made explicit (the one presently in 
mind is a very weak one) one may anticipate two surprising features: 
percolation at the critical point, 
and infinity of infinite clusters \cite{Aiz96}.   

\bigskip
\noindent{\bf Remark:} The above observation may alarm those familiar
with lattice percolation models, but the apparent conflict with 
the general uniqueness Theorems \cite{AKN,BK} is not a real 
contradiction; the general result of Burton--Keane \cite{BK} requires 
discreteness on some  short--scale.  

Our discussion of the scaling limit continues now in the generality of
of Type I models.  This covers the case of $2D$, and presumably 
applies also to dimensions $d = 3, 4, 5$ -- though  there are no 
rigorous results to support such claim.    We of course limit now
our attention to the critical regime.

\section{Formulation of the scaling limit; type I models} 

\subsection{The question}

The quantitative description of the continuum limit is expressed 
through a number of functions.  Two examples are described in 
Figure~\ref{F&G}.   The pictures drawn there refer to events defined 
on the macroscopic scale, with the probabilities considered in the 
limit $\alpha \to 0$.

\addtocounter{figure}{3}
\begin{figure}[ht]
    \begin{center}
    \leavevmode
    \epsfbox{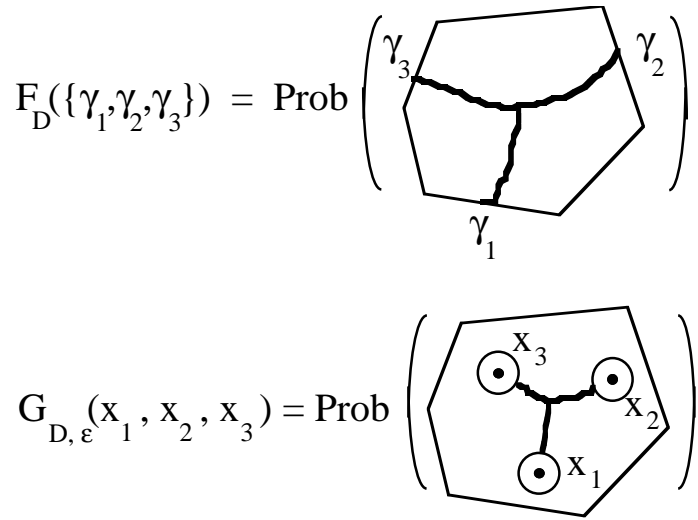}
  \caption{Two functions associated with the continuum limit 
  ($\alpha =$ {\em (lattice spacing)} $\to 0$) of Type I critical models:  
  1) $F$ is the limit of the probabilities that boundary segments are 
  connected as  indicated, 2) $G$ is the limit of probabilities
  that the neighborhoods of the points \{$x_1, x_2, x_3$\} are connected.}
\label{F&G}
\end{center}
\end{figure}

The question we shall address next is what mathematical
object would capture, in a natural way, those geometric features
of the Incipient Spanning Clusters which are visible in the scaling 
limit.  This can be rephrased as asking what stochastic--geometric 
object embedded in $R^d$ can be associated with the functions 
referred to above, and others of this kind, so that they can be 
naturally viewed as the connectivity probabilities (and not just 
limits of \ldots).  

In the absence of a direct insight a canonical approach could be to
to  define the object by the list of its quantifiers, and introduce on 
the space of those some minimal $\sigma$ - algebra which would allow 
to bring in probabilistic notions.  However, as the example of the 
Wiener process (Brownian motion) shows, it may be worthwhile to
learn the regularity properties of the geometric  object under 
consideration.

\subsection{The first attempt: ISC as a subset of $\Lambda$}\mbox{}

At first sight, it seems natural to regard the ISC as
{\em closed subsets} of the region $\Lambda \subset R^d$.  
However, this formulation will not serve our
purpose.

Before dismissing this attempt, let us recall that what makes 
the collection of closed subsets of  $\Lambda\subset R^d$
into a particularly convenient space for the description of 
random fractals (\cite{Fal}) is the  following classical result.
\begin{thm} For any compact metric space
$\Lambda$, the space \be
\H(\Lambda) = \{ A \subset \Lambda : \mbox{\rm \ A is closed } \}
\ee
is compact in the corresponding Hausdorff metric.
\end{thm}
The Hausdorff metric $h(A,B)$ is defined so that
$ h(A,B) \le \epsilon $ {\em if and only if}
$ dist(x,B) \le \epsilon$ for any $x \in A$, 
and $ dist(y,A) \le \epsilon$  for any $y \in B$.

Alas, the features which are of interest to us, such as the
existence and location of connecting paths,
are not continuous in the Hausdorff metric. 
Moreover, the configurations of critical percolation models are
typically among the points of discontinuity.  The reason is that in
typical configurations there are ``choke points'' where a  small-scale
change, possibly of a {\em single bond}, or {\em cell},
would drastically alter the available connecting paths (as indicated
in Figure~\ref{low-high}a, and more explicitly in 
Figure 1d of ref.~\cite{Aiz96}).
Such a change shifts the point in  $\H(\Lambda)$
only a distance of the order $O(\alpha)$, which is not detectable in
the scaling limit. However, the effect on the available connecting
paths is clearly visible on the large scale.

\subsection{H\"older continuity of the connecting paths} 

Since the scaling limit of the set does not capture
the information on the realized connections, it is natural
to include that information explicitly in the description of 
limit.  Some of it is expressed through the collection of the
realized self-avoiding paths, each given by
a continuous function $g: [0,1] \to \Lambda$.   
In the terminology of \cite{Sta_rev}, we are including both the 
{\em backbone} (BB)
paths and the paths connecting the {\em dangling ends} to the 
backbone.

Potential obstacles in describing the configuration through
the realized paths are:
\begin{itemize}
\item[1.] The possibility
that as the short scale is refined the connecting paths
could become increasingly irregular.  It is not a-priori
obvious that in the limit $\alpha \to 0$
the connections can still be
expressed through continuous functions.
(One could worry here about the need to 
consider more general continua [connected closed sets].  Their 
collection is somewhat unwieldy, e.g., some continua do
 not support the image of any continuous non-constant function.) 
\item[2.]  It is not initially clear whether the information
provided by the set of the connected paths suffices for
questions concerning higher order connections.
If not, then one might need to list also connected
line graphs of higher complexity.
\end{itemize}

The first concern is completely answered by the following
result \cite{Aiz96b,AizBur97} 
(see Note Added in Proof, next page).

\bigskip

\begin{thm} For any critical Type I percolation model
all the realized 
connected (self- avoiding) paths in a compact region
$\Lambda \subset R^n$ can be simultaneously
parameterized by uniformly continuous functions, 
$\underline{g}(t)\ \ 0 \le t \le 1$,  satisfying
the H\"older continuity condition:
\begin{equation}
\frac{|\underline{g}(t_1)-\underline{g}(t_2)|}
{|t_1-t_2|^{\lambda}  } \ \le \ \kappa(\omega) \ ,
 \ \ \mbox{for all \ 
$0\le t_1, t_2 \le 1$} \ ,
\end{equation}
with some fixed $ 0 < \lambda < 1/d$ and 
a configuration dependent continuity modulus  
for which
\bea
Prob_{p_c,\alpha}\left( \kappa(\omega) \ge t \right) \ & \le& \ g(t) 
\nonumber \\
& \rightarrow & 0 \ \ \ , \mbox{ for $t \to \infty$ }
\eea
uniformly in $\alpha$.
\end{thm}

In other words: in Type I models, in the critical regime one
seldom finds a connected path in $\Lambda$
which cannot be ``traced'' in a unit of time
by means of a ``fairly regular'' function.
The continuity condition we use is consistent even with 
a ``fractal''
landscape, and consequently the regularity does not deteriorate
as $\alpha \to 0$.  (That is not true for $p>p_c$.)

The {\em self--avoidance} condition is applied only on the microscopic 
scale, and should be interpreted in the sense which 
is natural for the model, e.g., for bond percolation the paths 
should not repeat any bond, and for the random Voronoi--tessellation 
the paths should not re--enter any cell.    
We note that the paths need not appear self--avoiding when 
viewed on the macroscopic scale (in the limit $\alpha \to 0$).

Concerning the second of the above reservations, the question has a
simple answer in $d=2$ dimensions, though the situation in higher
$d$ is still not as clearly resolved.  
The basic issue is: how to determine if a pair of paths which
on the scale of the continuum seem to intersect are actually connected
on the microscopic scale.  As we discussed, there are situations in which 
two connected paths come within distance $\alpha$ without touching
and without there being another path linking the two.
Conveniently, at least in $d=2$ dimensions such 
{\em close encounters 1st kind} do not occur at non-terminal points,
in the sense which is stated precisely in ref.~\cite{Aiz96}, and proven
in ref.~\cite{Aiz96b}.

\subsection{Our choice: ISC as a collection of realized H\"older-- \\ 
continuous paths} 
\footnote{{\bf Note Added in Proof:\/}   An equivalent but possibly
more appealing formulation is to describe ISC through the collection of 
the realized paths of finite tortuosity. This approach, along with some basic 
results concerning random curves with bounded
 tortuosity, is being
developed in a joint work with  Almut Burchard \cite{AizBur97}.}

To formulate the limiting representation of 
the Backbone and the Incipient Spanning
Clusters, we find it convenient to first  represent the percolation 
configuration  by the
random  collection of all the realized (connected) paths in
$\Lambda$ which are regular and self-avoiding in the sense explained above. 
That random object, which we call the {\em percolation Web} ($W$, or 
$W(\omega)$), is of the form 
\begin{equation}
W  \subset C_{1/d}([0,1], R^d)  \; ,
\end{equation}
where $C_{1/d}([0,1], R^d)$ is the space of H\"older--continuous 
functions:      
\bea 
C_{1/d}([0,1], R^d) = \left\{ g \in C([0,1], R^d) \ {\huge :}
\sup_{0\le t_1,t_2 \le 1} 
\frac{|\underline{g}(t_1)-\underline{g}(t_2)|}
{|t_1-t_2|^{1/d}  } < \infty 
\right\}  \nonumber \\
\mbox{ }
\eea
We shall also denote by $W_{\Lambda}(\omega)$ the restriction 
of $W(\omega)$ to functions with range in 
$\Lambda \subset \mathtt{R}^d$
(i.e. to $C([0,1], \Lambda)$).

In $d=2$ dimensions, the range of values of $W(\omega)$  is 
constrained by the following {\em consistency conditions},
of which the first three are obvious, but the
fourth one reflects a non-trivial fact (there [typically] 
are no close encounters of the $1$st kind,
in the limit $\alpha \to 0$). 
 
\bigskip
\noindent {\bf Percolation web consistency conditions} 
\begin{itemize}
\item [C1]  (Closure) $W_\Lambda(\omega)$ is closed as a subset 
of $C([0,1], \Lambda)$.   
\item [C2]  (Reparametrization--invariance)  
For each realized path $f \in W_\Lambda(\omega)$  
any path of the form $\tilde{f}(t)=f(\tau(t))$, with $\tau(\cdot)$
continuous and of bounded derivative [or just a Lipschitz function], 
is also realized.  
\item [C3]  (Splicing stability) 
If two paths of $W_\Lambda(\omega)$ intersect at non-terminal
points, then the  paths obtained by different ``splicings'' of 
the four resulting segments are also realized.   
\end{itemize}

We denote by $\Omega_{\Lambda}$ 
the collection of subsets of $C([0,1], \Lambda)$ satisfying the 
consistency conditions C1 -- C3.

(Trying not to be too formalistic here, let us just note
that $\Omega = \Omega_{R^d}$ is a complete separable metric space.
The continuum limit of critical percolation models in a macroscopic
region  $\Lambda$ will be 
described by probability measures on the natural $\sigma$
algebra on $\Omega(\Lambda)$.)

The {\em Backbone} can now be described by the
collections of paths in the Web which traverse $\Lambda$, 
\begin{equation}
B_\Lambda(\omega) = \left\{ f \in W_\Lambda(\omega) : 
g_1(0)=-\ell, \; \; g_1(1)=\ell \right\}
\end{equation}
and the collection of the Incipient Spanning Clusters 
is described by a collection of
{\em pairs of paths}, of the form:
\begin{equation}
S_\Lambda(\omega) = \left\{(f,g) {\Big |} 
\begin{array}{l}
  \mbox{$f \in W_\Lambda(\omega), \; \; g \in B(\omega)$,  \ with} \\
  \mbox{$\underline{f}(0) = \underline {g}(t)$ for some 
       and $0 \le t \le 1$ }
\end{array}  \right\}  \ \ ,
\end{equation}
where $(f,g) \in S_\Lambda(\omega)$ is taken to imply that there
is an actual contact between the paths (at the microscopic level,
which is otherwise no longer visible).

\section{The web --- an existence result}

{}From the perspective of the continuum limit, the microscopic 
model is a construction scaffold.
When it is removed, a more remarkable structure is 
exposed (as in Emily Dickinson's metaphor).  

Once we have a direct way to formulate the continuum theory, 
it is mathematically natural to restart the discussion 
and pose the question in the standard {\em existence} and
{\em uniqueness} terms.  It should be appreciated that a successful
formulation of the uniqueness question will shed  light
on the phenomenon of {\em universality} in critical behavior.

Following is an existence result \cite{Aiz96b}.

\bigskip

\begin{thm}  For each dimension in which the critical behavior is 
Type I, there is a one--parameter family of probability
measures ($\mu$) on $\Omega$
which have the following properties. 
\begin{itemize}
\item[1.] {\em (Independence)} For disjoint closed regions,  
$A \subset B \subset R^d$, $W_A(\omega)$ and $W_B(\omega)$
are independent [as random variables].     
\item[2.] {\em (Euclidean invariance)}  The probability measure is 
invariant under translations and rotations.   
\item[3.] {\em (Regularity)} 
The spanning probabilities of compact rectangular
regions are neither $0$ nor $1$:
\end{itemize}    
\begin{equation}
R_{s}:= Prob{[\mu]}\left(
\begin{array}{c}
        \mbox{there is in $[-s,s]^d$  a }  \\
        \mbox{left $\leftrightarrow$ right spanning cluster}
\end{array}  \right) \ \ 
\stackrel{\textstyle > \; 0}{_{_{\textstyle < \; 1} }} 
\end{equation}  
\label{thm3}
\end{thm}

A convenient parametrization within the family of measures is 
the crossing  probability $R_{1}$. 

The measures are constructed as continuum 
limits of sequences 
of models with suitably adjusted percolation densities.  If the 
standard picture is correct, the density needs be adjusted with the 
lattice spacing as:
\be
p \ = \ p_c + (R_1-1/2) \ Const. \ \alpha^{1/\nu} \ .
\ee
In order to end with a rotation invariant measure, we start from 
either the droplet percolation model, or the Voronoi-tessellation
percolation model.   

\bigskip 

\noindent {\bf Some open problems:}  
\nopagebreak 
\begin{description}
        \item[Convergence --]  A characteristic shortcoming of the available
methods is the lack of proof of  convergence of the scaling limit.  
Our construction relies on compactness arguments, 
which guarantee convergence along subsequences.  Proof of convergence
will be an outstanding technical contribution to the subject.
Alternatively stated, this is a question of uniqueness of the scaling
limit.

        \item[Uniqueness --]  
A broader formulation of the uniqueness question is:  \\ 
do the three 
conditions seen in the existence result, Theorem~\ref{thm3},
limit the range to only the one--parameter family of measures?  \\ 
If not, are there additional assumptions which would?  \\
Also: is the full rotation invariance of such measures implied by
just the  rotation invariance of the cube--crossing probabilities? 
\end{description}
Since the measures in question include all the continuum
limits of critical percolation models, positive answer would 
cast in a clear mathematical form  some of the expected 
universality of critical behavior -- in a sense which was clearly 
articulated only relatively recently, in Langlands et.al. 
\cite{LPPS} and related works 
[15, 6-9].

Other mathematical challenges are to affirm (or test) the 
Renormalization Group picture, which suggests some exact properties 
for the  constructed measures, and establish the conjectured 
conformal invariance of the critical measure (a special member
of the one-parameter family).  

\section{Relation with the renormalization group}

The continuum object constructed in Theorem~\ref{thm3} bears
an interesting relation to the  RG picture.
While still no sensible formalism has been found
for an {\em exact} representation of the renormalization group 
transformation as a map, 
the one--parameter family of measures presented in 
Theorem~\ref{thm3} may be viewed as corresponding exactly 
to what would be the unstable (i.e., expanding) fiber 
extending from the critical point in such a space.
Along this fiber, the RG maps coincide with dilatations.

\addtocounter{figure}{4}
\begin{figure}[ht]
    \begin{center}
    \leavevmode
    \epsfbox{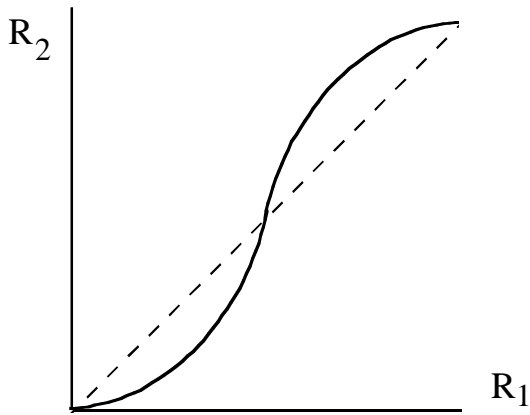}
  \caption{The conjectured relation between two crossing probabilities 
  in the one-parameter family of the constructed measures (schematically).  
  The map
  this function generates on $[0,1]$ corresponds to the action of the 
  Renormalization Group along the expanding fiber.  Assuming 
  the standard RG picture, the slope of the function at its middle 
  [unstable]
  fixed point is {\em exactly} $2^{1/\nu}$, 
  $\nu$ being the 
  correlation--length critical exponent.}
\label{RG}
\end{center}
\end{figure}

It is interesting to plot the joint values of
$R(s)$ and $R(1)$ (say, for $s=2$).
The standard  RG picture leads us to expect 
there the $S$-shape of a function which as a map of the unit 
interval into itself has  one unstable fixed point
and two stable fixed points at $0$ and $1$ (as in Figure~\ref{RG}).  
The slope of the function at the unstable fixed point 
should be {\em exactly} $s^{1/\nu}$.
(For $2D$ the predicted value is $\nu = 4/3$, den Nijs \cite{Nij}). 

Even in the absence of an exact setup, the renormalization group 
picture has provided very effective approximate tools \cite{RSK}.
One may view Figure~\ref{RG} as presenting a  limit of the 
``cell to cell''
renormalization group map, which recently became appreciated as 
one of the more effective approximate RG methods 
\cite{Z,Hu_rg,HCW}.

\section{Fractal structure}\label{fractal}

In Type I models, the number of clusters connecting the boundary of a 
cube $\Lambda \subset R^d$, centered at $0$,  with the contracted
cube $s \Lambda$, $0<s<1$, is finite; in the  sense that the 
probability for observing $k$ such clusters obeys bounds which 
are uniform in the 
short--distance/lattice--spacing $\alpha$ (as $\alpha \to 0$),
and decay to zero for $k \to \infty$.  
Correspondingly, the scaling limit exhibits (a.s.) only a 
finite number of such clusters 
(once one knows how to count them), 
and altogether only  countably many macroscopic size clusters
in $\Lambda$ (the infinity is caused by the union over all scales). 
Furthermore, 
the number of ``left $\leftrightarrow$ right'' spanning clusters 
remains finite in the continuum limit.  

Details of the expected fractal structure were discussed in 
the review article of Stanley \cite{Sta_rev} in the context of 
lattice models.  New considerations are added when one looks at 
the scaling limit.  To present some basic results, 
let us start with--

\noindent{\bf Definition:} 
{\em 1.   For a given configuration of the percolation Web, 
$W_{\Lambda}(\omega)$, 
we say that two sites, $x,y \in \Lambda$ are {\em connected} if 
$W_{\Lambda}(\omega)$ includes a path visiting each of them.  

\noindent 2.  The connected cluster of a site $x \in \Lambda$
is the union of the sites $y \in \Lambda$ connected to it.  We denote 
it $\C(x)= \C_{\omega}(x)$.

\noindent 3.  $\C(\partial \Lambda)$ is the collection of sites 
connected (as in 1.) to the boundary $\partial \Lambda$.

\noindent 4.  The ramification number $R_{\omega}(x)$ is the 
maximal value of $k$ for which there are $k$ 
paths in $W(\omega)$ starting at $x$ and otherwise non-intersecting.}

There are some surprises:  first is the lack of 
transitivity of the relation ``$x$ is connected to $y$''.  We view 
this not as a shortcoming of the terminology, but rather as an 
expression of an interesting phenomenon, related to the existence of 
tenuous connections (when a pivotal bond is reversed, one obtains a 
configuration in which two realized paths meet without 
being connected on the microscopic level, an example is indicated in 
Figure~\ref{low-high}a).  

The second ``surprise'' is the first statement in the following 
list of properties of the scaling limit, ref. \cite{Aiz96b}. 

\bigskip

\begin{thm}  For Type I critical models, in typical realizations of 
the percolation Web:

1.  The connected clusters of (Lebesgue-\/){\em almost all} 
sites $x \in \Lambda$ contain no other site, i.e. $R(x)=0$ or
\be
\C(x) = \{x\} \ .
\label{cx}
\ee

2.  The collection of sites violating \eq{cx}, which includes the 
random set $\C_{\omega}(\partial \lambda)$, is of Hausdorff dimension 
$\le (d - \lambda)$ (as defined by 
\eq{lambda}.   Furthermore,
 
3. The above set is of finite ramification; there is a non-random value
$k < \infty$ such that  $R_{\omega}(x) \le k$ for {\em all} (not just a.e.) 
$x\in \Lambda$.
\end{thm}

For $2D$, we guess that the maximal ramification number is about $k=5$,
though that is still not fully resolved.  

\section{Conjectured conformal invariance} \label{Conf-inv}

The particular measure which corresponds to the fixed--point
value of $R_1$ is expected to be {\em fully} dilatation invariant.
It is conjectured that it is also strictly covariant under 
{\em conformal maps}.  

\addtocounter{figure}{5}
\begin{figure}[ht]
    \begin{center}
    \leavevmode
    \epsfbox{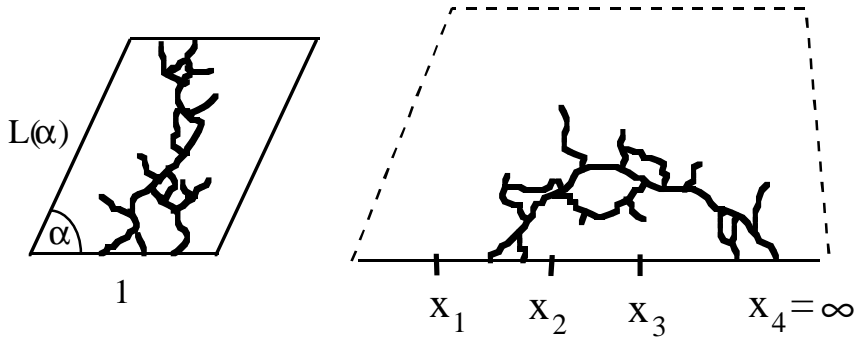}
  \caption{The setup for the question of Langlands et al. \protect\cite{LPPS}
  and Cardy's surmise.}
\protect\label{conformal}
\end{center}
\end{figure}

The conformal invariance is expected in any dimension, but
the conformal group is particularly large in $2D$ and it is there
that the related considerations have been shown to have powerful 
consequences, and have led to explicit predictions
concerning critical behavior in a rich collection of models 
\cite{BPZ,Gin}.  For the rest of this section we 
restrict the attention to two dimensional systems.

The $2D$ scaling limit was studied numerically by Langlands et.al. \\ 
(LPPS) \cite{LPPS}.  In addition to testing the universality of
the spanning probabilities (in a sense which broke new grounds 
\cite{Z,AH94,SAA,HA96}),  
LPPS asked how should the aspect ratio of a parallelogram be adjusted
with the angle (see Figure~\ref{conformal}), if one wants to keep 
the spanning probability constant.  This author's suggestion that the 
criterion should be conformal equivalence fitted well with the 
numerical data.   A much more complete answer was proposed by 
J. Cardy \cite{Car}, who produced a differential equation for the 
upper--half--plane version of the problem (to which the original one is 
reduced via the corresponding Riemann map).  Cardy's equation for the 
crossing probability has a unique solution with the natural boundary
conditions, and the resulting function of the angle
$\alpha$ and the length $L$  (see Figure~\ref{conformal})
was found to be in perfect agreement with the numerical 
results of LPPS. 

Cardy's equation drew on a field--theoretic perspective, 
and on an extrapolation based on relations with some other models of 
Statistical Mechanics.  Other insights followed, including Pinson's 
proposal \cite{Pin} for the exact values of the 
probabilities of different windings for twisted boundary conditions. 
The reader is referred to \cite{LPS} for an account of these 
developments.   

The bottom--up derivation of the conformal invariance is still an 
open challenge.  There are some partial results, such as the following
one (whose proof is not that difficult) which is conditioned on
a strong assumption about the scaling limit.   

The result stated below follows \cite{Aiz96b}, but 
we note that a related statement with a somewhat different formulation 
was presented in  \cite{BS} ([the two were arrived at independently]).
To formulate the proven assertion, let us first present a statement 
which upon some consideration appears believable, though its proof has
still eluded us. 

\bigskip
\noindent {\bf Conjecture:} 
{\em The Voronoi-tessellation percolation models with 
position--dependent density profiles of the form 
$\rho(x) = t\cdot g(x)$, with $g( \cdot )$ continuous 
and non--vanishing in  $\Lambda \subset R^2$, the limit 
$t \to \infty$ exist (for the ISC and for the Web processes),
and is independent of the function $g( \cdot )$.}

\bigskip
\begin{thm} 
In $d=2$ dimensions, under the above conjecture and assuming also the 
critical behavior is Type I, the scaling limit of the Voronoi 
percolation model is conformally invariant, in the 
sense that for any map $T\ : \ \Lambda \rightarrow R^2$ which is 
invertible and conformal on $\Lambda$, the image of the ISC [Web]
process
in $\Lambda$ coincides with the ISC [Web] process in $T\Lambda$.
\end{thm}

A particular expression of the conformal invariance is that the 
function defined in Figure~\ref{F&G} satisfies
\be
F_{D}(\{\gamma_1,\ldots,\gamma_k\})\ = 
\ F_{TD}(\{\gamma_1,\ldots,\gamma_k\})  \  ,
\ee
where $\{\gamma_1,\ldots,\gamma_k\}$ refers to a collection of boundary 
segments which are to be connected by a common cluster.  (One may 
of course generate a large number of other, similar, functions.)

One may employ also the Riemann map which takes the interior of $D$
conformally onto the upper--half--plane; Figure~\ref{conformal} 
indicates two such events for which conformal 
invariance implies equality of probability.  
In case the boundary of $D$ consists of four segments, conformal 
invariance implies that the probability 
depends only on one cross--ratio, i.e.,
is given by a function of the form 
\be
f(x_1,\ldots,x_4) = \phi(\ u(x_1,\ldots,x_4) \ ) \ ,
\label{u}
\ee
with
\be
u(x_1,\ldots,x_4) \ = \ 
\frac{(x_1 - x_2 ) (x_3 - x_4)}{(x_1 - x_3 ) (x_2 - x_4)} \  .
\ee

\section{Cardy's equation}

The equation which Cardy proposed for the above quantity
can be transcribed as \cite{Car,LPS}:
\be
u(1-u) \frac{d ^2 \phi}{d u\mbox{ }^2} +
 \frac{2}{3} (1-2u) \frac{d  \phi}{d u} \ = \ 0  \ ,
\ee
with the boundary values $\phi(0) = 0$, $\phi(1) = 1$.  
The solution can be presented in the integral form:
\be
\phi(u) \ = \ \int_{0}^{u} \frac{d\ x}{[x(1-x)]^{2/3}} \  / \ Norm.
\label{phi}
\ee

Some algebraic aspects of Cardy's equation are more visible 
when it is expressed in terms of
the differential operators:
\be
\L_n \ = 
\ - \sum_{j\ne 1}\left( x_j - x_1 \right)^{n+1} 
\frac{\partial}{\partial x_j }   \   .
\ee
In terms of these, the equation (as originally presented)  is
\be
\left( \L_{-2} - {\textstyle \frac{3}{2}} \L_{-1}^2 \right) 
f(x_1,\ldots,x_4) \ = \ 0  \  .
\label{cardy}
\ee

Cardy's original argument \cite{Car} invoked an analogy with  
equations describing the effects on the free energy of 
certain Potts models of changes in the boundary 
conditions.  
Various step in this approach are still beyond the reach of 
rigorous methods.  Let us, however, point out that the very 
intuitive hypothesis of invariance under 
conformal maps which preserve the relevant domain
(plus the necessary differentiability) implies:
\be
\L_1 \left( \L_{-2} - {\textstyle \frac{3}{2}} \L_{-1}^2 \right) 
f(x_1,\ldots,x_4) \ = \ 0  \  .
\label{precardy}
\ee
The derivation of this equation 
is an elementary and amusing 
exercise in the Virasoro algebra, of the
commutation relations:
\be
\left[ \L_n,\L_m \right] \ = \ (n-m) \ \L_{n+m}  \  ,
\label{virasoro}
\ee
starting from the observation that the conformal invariance 
assumption implies
\be
\L_0 f(\ldots) \ = \ 0\ \ , \mbox{and} \  \ \L_1 f(\ldots)\ = \ 0
\ .
\label{Lo}
\ee
Equation (\ref{Lo}) holds since the two operators generate flows 
preserving $f(\ldots)$, being 
associated with M\"obius transformations which preserve the
upper--half--plane and the point $x_1$
($\L_{-1}$, however, corresponds to a shift of $x_2,\ldots$ which 
leaves $x_1$ behind).
Thus, when the two factors in \eq{precardy} are 
transposed $f(\ldots)$ is annihilated by $\L_1$.  
The commutator of the two 
terms is a combination of $\L_0$, $\L_{1}$ and a term proportional
to $\L_{-1}$ which is eliminated through the judicious choice of the
coefficient $3/2$ in \eq{precardy}.  

The transition from \eq{precardy} to \eq{cardy} amounts to 
the  removal of the former's leftmost factor. 
That is however a big step, since it
 transforms an equation with limited content into one 
which completely determines the solution.  (Although one should 
not be too dismissive of \eq{precardy}: in terms of the conformal 
invariance structures, it conveys the fact that for the percolation 
problem the {\em conformal charge} is zero, which is a speculation
born out by the Russo--Seymour--Welsh theory \cite{R,SW}.)  

Though we are  still short of a proof, it turns out that one may 
explain  from the
Continuum Percolation Web's perspective a mechanism for the 
extraction of Cardy's equation, \eq{cardy}, from \eq{precardy}.
The argument 
employs a plausible description of an effect caused by 
the separation of scales (\cite{Aiz96b}).

\section{Selected characteristic exponents}

\addtocounter{figure}{6}
\begin{figure}[ht]
    \begin{center}
    \leavevmode
    \epsfbox{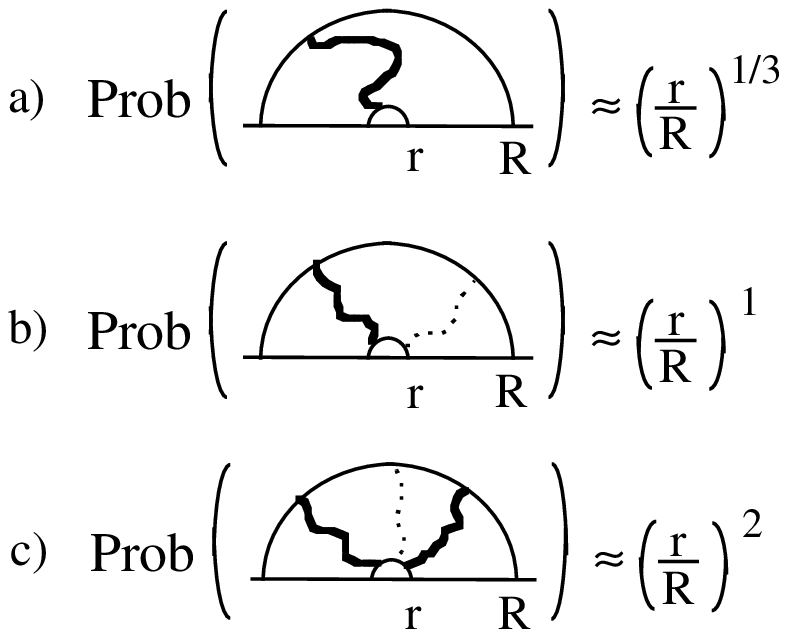}
  \caption{Some characteristic exponents associated with spanning 
  probabilities in $2D$.  Solid lines indicate spanning clusters, 
  and dotted lines indicate spanning dual, separating, clusters. 
  Example a) is implied by Cardy's equation, b) and c) are based
  on more elementary observations, \protect\cite{K_scaling,Aiz96b}. }
\protect\label{exponents}
\end{center}
\end{figure}

An interesting aspect of the explicit solution, \eq{phi},
is that it yields
the power $\epsilon^{1/3}$ for the 
probability that a short interval, of size $\epsilon$, 
is connected in the half plane a distance of order $1$ away.
This exponent does not appear to be obvious to the naked eye.
In Figure~\ref{exponents} we present it next to two other 
characteristic exponents which have a more elementary 
derivation (discussed along with other examples and applications
in ref.~\cite{K_scaling,A_Web,Aiz96b}).

Less one would be lulled by the simplicity of the exponents seen 
in Figure~\ref{exponents}, let us mention that the probability 
of a full annulus to be spanned is predicted to 
behave as $(r/R)^{5/48}$, den Nijs \cite{Nij}.

\noindent{\bf Question:}  What are the values of the 
characteristic exponents for $k$ disjoint spanning clusters,
in the full or cut annulus, for all the other values of $k$?  

The exact answer may be within the reach
of the methods of Conformal Field Theory \cite{Gin,Car}, 
or the Coulomb--gas representation, \cite{FSZ}, 
which has been recently discussed by T. Spencer\cite{Spe}.  
It was recently proven by other methods that for large $k$ the 
exponent is of the order of $Const. \ k^2$, \cite{Aiz96}.   

\section{Relation with field theory}

The continuum object described here in earlier sections
is related to a number of ``field theories''.   Cardy's 
equation for the function $F$ drew on its analogy with 
the vacuum expectation value of a product of 
operators switching the boundary conditions (in Potts models).
Another field theoretical object is related with the 
function $G$, also defined in Figure~\ref{F&G}.  Its simplest
manifestation is in the {\em one--point function} defined next.

In the scaling limit of a Type I model, the probability that
a given site $x\in D$ is
connected to the boundary of the domain $D$ is zero, 
but the probability that a small ball around it, 
$B(x, \epsilon) = $ \mbox{ $\{ y \ : \ |y-x| \le \epsilon \} $},
is connected to $\partial D$ is positive.  
Based on the ideas mentioned in Section 5, one expects that for a 
suitable $d_w$ the following limit exists
\be
\ \ \ \  h_{D}(x) \ = \ \lim_{\epsilon \to 0}\  \epsilon^{-k (2-d_w)}\ 
Prob\left(  
\mbox{$B(x, \epsilon)$ is connected to $\partial D$} 
          \right) \ .  
\label{def_h}
\ee
If so, and if the probability measure maps covariantly under 
conformal maps,
then, for any conformal transformation $T\ : D\to \mathtt{R}^d$
\be
h_{D}(x) \ = \ (|\det(\partial T/ \partial x )|^{1/d})^{2-d_w} 
\ h_{TD}(Tx) \  .
\ee

The function $h_{D}(x)$ can be viewed as the expectation value of
an entity defined by the limit
\be
\psi_{D}(z) \ = \ \texttt{w}-\lim_{\epsilon \to 0} \ \epsilon^{-(2-d_w)} 
\ \mathtt{I}[ \mbox{$B(x, \epsilon)$ is connected to $\partial D$}] \ .
\ee
The limit is initially interpreted in the weak sense (i.e., 
expectation values of \ldots).  If we try to give $\psi_{D}(z)$ 
a stronger meaning, in the context of the continuum process discussed 
in earlier sections, we find that for a given realization of the 
continuum percolation Web, $\psi_{D}(z)$ is zero at almost all points,
$\{ z \in D\}$,
but it diverges along the fractal set connected to the boundary.  It 
would be natural to think of it as a distribution--valued random
field, but clearly some more thought should be given for a complete
development of this interpretation.  Related concepts on the horizon 
are:  the {\em stress--energy tensor} for percolation, and 
{\em operator product expansions} for entities like $\psi_{D}(z)$ 
(discussed in other contexts in \cite{KC,BPZ,Car_rev,Gin}). 

Let us conclude by noting that similar considerations apply to the
{\em n--point function} associated with the function $G_{D,\epsilon}$ 
defined in Figure~\ref{F&G}.  

If, as in \eq{def_h},
the following limit exists, 
\be
g_{D}(z_1, \ldots, z_n) \ = 
  \  \lim_{\epsilon \to 0} \epsilon^{-k (2-d_w)}
  G_{D,\epsilon}(z_1, \ldots, z_n) \  ,
\ee
then under the ``plain'' conformal invariance hypothesis 
the resulting function should satisfy
\be
g_{D}(z_1, \ldots, z_n) \ = \ 
\prod_{j=1}^{n}|T'(z_{j})|^{2-d_w} \ g_{TD}(Tz_1, \ldots, Tz_n) \  .
\label{g_law}
\ee
We switched here to notation appropriate for $2D$, where a rich 
class of conformal transformations is provided by the maps 
$T\ : D \to \mathtt{C}$ which are analytic and invertible on $D$. 
Equation (\ref{g_law}) is reminiscent of the transformation law of the 
vacuum expectation values of products of field operators 
in conformal field theory \cite{BPZ}, although the higher order 
connectivity function were not yet transcribed into such expectations of 
products of local field operators, and percolation seems to 
be missing the reflection positivity (corresponding to the 
positivity of the inner product) which plays an important role 
in field theories.    

It would be interesting to 
see further development of the field theoretic content of the 
continuum theory described here.  This should be done both within the
perspective native to percolation models, and in the direction of 
links with other field theoretic structures.  Existence of such 
links is suggested by the  Fortuin--Kasteleyn relations of percolation 
with Ising and Potts models \cite{FK}, and the relation of the 
latter with $\phi^{4}$ and other field theories.

\newpage 

\section*{Acknowledgments}
I wish to thank Kenneth Golden and Ohad Levy for informative discussions
of resistor networks and high contrast composites utilizing critical percolating 
structures, and Robert Langlands for stimulating discussions concerning the 
renormalization group map in the percolation context.  The organizers 
are thanked and congratulated for a very stimulating workshop at the IMA.  
This work was supported in part by the NSF grant PHY-9512729.


\begin{thebibliography}{10}

\bibitem{BI}
D.~J. Bergman and Y.~Imry, ``Critical behavior of the complex dielectric
  constant near the percolation threshold of a heterogeneous material,'' {\em
  Phys. Rev. Lett.},  {\bf 39},  1222  (1977).

\bibitem{composits}
D.~McLachlan, M.~Blaszkiewicz, and R.~Newnham, ``Electrical resistivity of
  composites,'' {\em J. Am. Ceram. Soc.},  {\bf 73},  2187  (1990).

\bibitem{binaries}
J.~P. Clerc, G.~Giraud, J.~M. Laugier, and J.~M. Luck, ``The electrical
  conductivity of binary disordered systems, percolation clusters, fractals,
  and related models,'' {\em Adv. Phys.},  {\bf 39},  191  (1990).

\bibitem{Sta_rev}
H.~E. Stanley, ``Fractal and multifractal approaches to percolation: some exact
  and not--so--exact results,'' in {\em Percolation Theory and Ergodic Theory
  of Infinite Particle Systems} (H.~Kesten, ed.), Springer -- Verlag, 1987.

\bibitem{LPPS}
R.~Langlands, C.~Pichet, P.~Pouiliot, and Y.~Saint-Aubin, ``On the universality
  of crossing probabilities in two-dimensional percolation,'' {\em J. Stat.
  Phys.},  {\bf 67},  553  (1992).

\bibitem{Z}
R.~Ziff, ``On the spanning probability in 2{D} percolation,'' {\em
Phys. Rev. Lett.},  {\bf 69},  2670  (1992).

\bibitem{AH94}
A.~Aharony and J.-P. Hovi, ``Comment on `{S}panning probability in 2{D} 
percolation'~''.  {\em Phys. Rev. Lett.}, {\bf 72}, 1941 (1994).

\bibitem{SAA}
D.~Stauffer, J.~Adler, and A.~Aharony, ``Universality at the three-dimensional
  percolation threshold,'' {\em J. Phys. A},  {\bf 27},  L 475  (1994).

\bibitem{HA96} 
J.-P. Hovi and A.~Aharony, ``Scaling and universality in the spanning 
probability for percolation'', {\em Phys. Rev. E}, {\bf 53}, 235 
(1996).

\bibitem{A_Web}
M.~Aizenman, ``The {C}ritical {P}ercolation {W}eb: Construction and conjectured
  conformal invariance properties,'' in {\em \em STATPHYS 19, Proceedings
  Xiamen 1995} (H.~Bai-lin, ed.), World Scientific, 1995.

\bibitem{Hu}
C.-K. Hu and C.-Y. Lin {\em Phys. Rev. Lett.},  {\bf 77},  8  (1996).

\bibitem{Aiz96}
M.~Aizenman, ``On the number of incipient spanning clusters.''
\newblock Nuclear Physics B [FS] {\bf 485}, 551 (1997).

\bibitem{Sen}
P.~Sen, ``Non-uniqueness of spanning clusters in 2 to 5 dimensions,'' {\em Int.
  J. Mod. Phys. C},  {\bf 7},  603  (1996).

\bibitem{Man}
B.~Mandelbrot, ``Fractals in physics: Squig clusters, diffusions, fractal
  measures, and unicity of fractal dimensionality,'' {\em J. Stat. Phys.},
  {\bf 34},  895  (1984).

\bibitem{Car}
J.~Cardy, ``Critical percolation in finite geometries,'' {\em J. Phys. A},
  {\bf 25},  L201  (1992).

\bibitem{LPS}
R.~Langlands, P.~Pouiliot, and Y.~Saint-Aubin, ``Conformal invariance in
  two-dimensional percolation,'' {\em Bull. AMS},  {\bf 30},  1  (1994).

\bibitem{K_IIC}
H.~Kesten, ``The incipient infinite cluster in two-dimensional percolation,''
  {\em Prob. Th. Rel. Fields},  {\bf 73},  369  (1986).

\bibitem{HASM}
J.-P.Hovi, A.Aharony, D.Stauffer, and B.B.Mandelbrot, ``Gap independence and
  lacunarity in percolation clusters,'' {\em Phys. Rev. Lett.},  {\bf 77},
  877  (1996).

\bibitem{CCD}
J.~Chayes, L.~Chayes, and R.~Durrett, ``Inhomogeneous percolation problems and
  incipient infinite cluster,'' {\em J. Phys A: Math. Gen.},  {\bf 20},  
  1521   (1987).

\bibitem{k_pc}
H.~Kesten, ``The critical probability of the bond percolation on the square
  lattice equals $1/2$,'' {\em Commun. Math. Phys.},  {\bf 74},  41  (1980).

\newpage   
\bibitem{K_scaling}
H.~Kesten, ``Scaling relations for percolation,'' {\em Commun. Math. Phys.},
  {\bf 109},  109  (1987).

\bibitem{ACCFR}
M.~Aizenman, J.~Chayes, L.~Chayes, J.~Fr{\"o}hlich, and L.~Russo, ``On a sharp
  transition from {A}rea {L}aw to {P}erimeter {L}aw in a system of random
  surfaces,'' {\em Commun. Math. Phys.},  {\bf 92},  19  (1983).

\bibitem{BS}
I.~Benjamini and O.~Schramm, ``Conformal invariance and {V}oronoi
  percolation.''
\newblock 1996 preprint.

\bibitem{Alex_RSW}
K.~Alexander, ``The {RSW} theorem for continuum percolation and the {CLT} for
  {E}uclidean minimal spanning trees.''
\newblock to appear in Ann. Prob.

\bibitem{AB}
M.~Aizenman and D.~Barsky, ``Sharpness of the phase transition in percolation
  models,'' {\em Commun. Math. Phys.},  {\bf 108},  489  (1987).

\bibitem{Men}
M.~Menshikov and A.~Sidorenko, ``Coincidence of critical points for {P}oisson
  percolation models,'' {\em Th. Prob. Appl.},  {\bf 32},  603 (in Russian; 547
  in translation)  (1987).

\bibitem{BGN}
D.~Barsky, G.~Grimmett, and C.~Newman, ``Percolation in half--spaces; equality
  of critical densities and continuity of the percolation probability,'' {\em
  Prob. Th. Rel. Fields},  {\bf 90},  111  (1991).

\bibitem{GM}
G.~Grimmett and J.~Marstrand, ``The supercritical phase of percolation is well
  behaved,'' {\em Proc. R. Soc. Lond. Ser. A},  {\bf 430},  439  (1990).

\bibitem{BCKS}
C.~Borgs, J.~Chayes, H.~Kesten, and J.~Spencer, ``The birth of the infinite
  cluster: finite--size scaling in percolation.'' In preparation.

\bibitem{R}
L.~Russo, ``A note on percolation,'' {\em Zeit. Wahr.},  {\bf 43},  39  (1978).

\bibitem{SW}
P.~Seymour and D.~Welsh, ``Percolation probabilities on the square lattice,''
  in {\em {\em Advances in Graph Theory} Annals of Discrete Mathematics}
  (B.~Bollob{\'a}s, ed.), vol.~3, North Holland, 1978.

\bibitem{FHS}
S.~Feng, B.~Halperin, and P.~Sen {\em Phys. Rev. B},  {\bf 35},  197  (1987).

\bibitem{Tou}
G.~Toulouse, ``Perspectives from the theory of phase transitions,'' {\em Nuovo
  Cimento B},  {\bf 23},  234  (1974).

\bibitem{HL}
A.~Harris, T.~Lubensky, W.~Holcomb, and C.~Dasgupta, ``Renormalization -- group
  approach to percolation problems,'' {\em Phys. Rev. Lett.},  {\bf 35},  327,
  (1975).

\bibitem{AN}
M.~Aizenman and C.~Newman, ``Tree graph inequalities and critical behavior in
  percolation models,'' {\em J. Stat. Phys.},  {\bf 36},  107  (1984).

\bibitem{HS}
T.~Hara and G.~Slade, ``Mean-field critical behavior for percolation in high
  dimensions,'' {\em Commun. Math. Phys.},  {\bf 128},  333  (1990).

\bibitem{BA}
D.~Barsky and M.~Aizenman, ``Percolation critical exponents under the triangle
  condition,'' {\em Ann. Prob.},  {\bf 19},  1520  (1991).

\bibitem{G}
G.~Grimmett, {\em Percolation}.
\newblock Springer-Verlag, 1989.

\bibitem{Sl}
G.~Slade, ``The lace expansion and the upper critical dimension for
  percolation,'' in {\em {\em Mathematics of Random Media} Lectures in Applied
  Mathematics}, vol.~27, Amer. Math. Soc., 1991.

\bibitem{coniglio}
A.~Coniglio, Shapes, surfaces, and interfaces in percolation clusters,
{\it in} Proc. Les Houches Conf. on Physics of finely divided matter
ed. M.~Daoud and N.~Boccara (Springer, Berlin, 1985).

\bibitem{AGK}
A.~Aharony, Y.~Gefen, and A.~Kapitulnik, ``Scaling at the percolation threshold
  above six dimensions,'' {\em J. Phys. A},  {\bf 17},  L 197  (1984).

\bibitem{AGNW}
S.~Alexander, G.~Grest, H.~Nakanishi, and T.~Witten, Jr., ``Branched polymer
  approach to the structure of lattice animals and percolation clusters,'' {\em
  J. Phys. A},  {\bf 17},  L 185  (1984).

\bibitem{DS}
E.~Derbez and G.~Slade, ``Lattice trees and {S}uper-{B}rownian {M}otion.'' 1996
  preprint.

\bibitem{Al}
D.~Aldous, ``Tree-based models for random distribution of mass,'' {\em J. Stat.
  Phys.},  {\bf 73},  625  (1993).

\bibitem{AKN}
M.~Aizenman, H.~Kesten, and C.~M. Newman, ``Uniqueness of the infinite cluster
  and continuity of connectivity functions for short- and long- range
  percolation,'' {\em Commun. Math. Phys.},  {\bf 111},  505  (1987).

\bibitem{BK}
R.~M. Burton and M.~Keane, ``Density and uniqueness in percolation,'' {\em
  Commun. Math. Phys.},  {\bf 121},  501  (1989).

\bibitem{Fal}
K.~Falconer, {\em Fractal Geometry}.
\newblock J. Wiley, 1990.

\bibitem{Aiz96b}
M.~Aizenman.  ``Scaling Limits for Percolation Models''
\newblock In preparation, title tentative.

\bibitem{AizBur97} 
M.~Aizenman and A.~Burchard,  ``Tortuosity Bounds for Random Curves'',
\newblock  In preparation.

\bibitem{Nij}
M.~den Nijs, ``Extended scaling relations for the magnetic critical exponents
  of the {P}otts model,'' {\em Phys. Rev. B},  {\bf 27},  1674  (1983).

\bibitem{RSK}
P.~J. Reynolds, H.~E. Stanley, and W.~Klein, ``A large-cell {M}onte {C}arlo
  renormalization group for {P}ercolation,'' {\em Phys. Rev. B},  {\bf 21},
  1223  (1980).

\bibitem{Hu_rg}
C.-K. Hu, ``Histogram {M}onte {C}arlo renormalization-- group method for
  percolation,'' {\em Phys. Rev. B},  {\bf 46},  6592  (1992).

\bibitem{HCW}
C.-K. Hu, C.~Chen, and F.~Wu, ``Histogram {M}onte {C}arlo position--space
  renormalization group: Applications to the site percolation,'' {\em J. Stat.
  Phys.},  {\bf 82},  1199  (1996).

\bibitem{BPZ}
A.~Belavin, A.~Polyakov, and A.~Zamolodchikov, ``Infinite conformal symmetry in
  two-dimensional quantum field theory,'' {\em Nucl. Phys. B},  {\bf 241},
  333  (1984).

\bibitem{Gin}
P.~Ginsparg, ``Applied conformal field theory,'' in {\em Fields, Strings and
  Critical Phenomena} (E.~Br{\'{e}}zin and J.~Zinn-Justin, eds.),
  North-Holland, Amsterdam, 1990.

\bibitem{Pin}
H.~Pinson, ``Critical percolation on the {T}orus,'' {\em J. Stat. Phys.},  {\bf
  75},  1167  (1994).

\bibitem{FSZ}
P.~di~Francesco, H.~Saluer, and J.~Zuber, ``Relations between the {C}oulomb gas
  picture and conformal invariance of two dimensional critical models,'' {\em
  J. Stat. Phys.},  {\bf 49},  57  (1987).

\bibitem{Spe}
T.~Spencer, private communication.

\bibitem{KC}
L.~Kadanoff and H.~Ceva {\em Phys. Rev. B},  {\bf 3},  3918  (1971).

\bibitem{Car_rev}
J.~Cardy, ``Conformal invariance and statistical mechanics,'' in {\em Fields,
  Strings and Critical Phenomena} (E.~Br{\'{e}}zin and J.~Zinn-Justin, eds.),
  North-Holland, Amsterdam, 1990.

\bibitem{FK}
C.~Fortuin and P.~Kasteleyn, ``On the random cluster model. {I}. introduction
  and relation to other models,'' {\em Physica},  {\bf 57},  536  (1972).

\end{thebibliography}

\end{document}